\documentclass[preprintnumbers,superscriptaddress,showpacs,pra]{revtex4}%
\usepackage[colorlinks=true,linkcolor=blue]{hyperref}
\usepackage{amssymb}
\usepackage{amsfonts}
\usepackage{amsmath}
\usepackage{graphicx}%
\providecommand{\U}[1]{\protect\rule{.1in}{.1in}}

\let\stdsection\section
\renewcommand\section{\nopagebreak\stdsection}
\begin{document}
\preprint{ }
\title{On relation between geometric momentum and annihilation operators on a
two-dimensional sphere}
\author{Q. H. Liu$^{a}$, }
\thanks{quanhuiliu@gmail.com}
\author{Y. Shen,$^{a,b}$ D. M. Xun,$^{a}$ X. Wang,$^{a}$ }
\thanks{hudawangxin@gmail.com}
\affiliation{$^{a}$School for Theoretical Physics, and Department of Applied Physics, Hunan
University, Changsha, 410082, China}
\affiliation{$^{b}$Department of Physics, 1102 IPST, University of Maryland, College Park,
MD 20742}
\date{\today }

\begin{abstract}
With a recently introduced \textit{geometric momentum} that depends on the
extrinsic curvature and offers a proper description of momentum on
two-dimensional sphere, we show that the annihilation operators whose
eigenstates are coherent states on the sphere take the expected form
$\alpha\mathbf{x}+i\beta\mathbf{p}$, where $\alpha$ and $\beta$ are two
operators that depend on the angular momentum and $\mathbf{x}$ and
$\mathbf{p}$ are the position and the \textit{geometric momentum,}
respectively. Since the geometric momentum is manifestly a consequence of
embedding the two-dimensional sphere in the three-dimensional flat space, the
coherent states reflects some aspects beyond the intrinsic geometry of the surfaces.

\end{abstract}

\maketitle

The coherent states on two-dimensional sphere, generated by the annihilation
operators, were discovered around the turn of present century, independently
by Hall \cite{H1,H3,H2,H4,H7,H5} in the Bargmann representation, and by
Kowalski and Rembieli\'{n}ski \cite{KR1,KR2,KR3,KR4} in the position
representation, respectively. Once each group of them got to know the work of
the other, both soon realized that their coherent states are essentially the
same \cite{H5,KR3}, and the equivalence was also noted by other group
\cite{VB}. However, it is puzzling that in the annihilation operators they
introduced, there is a fundamental quantity that is represented by a
non-hermitian operator and has the same dimension of linear momentum, but it
does not bear a transparent physical nor geometric meaning. This article
points out that the physical\ and geometric interpretation of the fundamental
quantity is easily available, based on the \textit{geometric momentum} that is
recently introduced to offer a proper description of momentum on sphere
\cite{Liu1}, whose general form for an arbitrary two-dimensional curved
surface with $M$ denoting the mean curvature, is given by
\cite{Liu1,Liu2,Liu3},
\begin{equation}
\mathbf{p}\equiv-i\hbar(\mathbf{r}^{\mu}\partial_{\mu}+M\mathbf{n}),\label{GM}%
\end{equation}
where $\mathbf{r=[}x(x^{\mu},x^{\upsilon}),y(x^{\mu},x^{\upsilon}),z(x^{\mu
},x^{\upsilon})\mathbf{]}$ is the position vector on the surface,
$\mathbf{r}^{\mu}=g^{\mu\nu}\mathbf{r}_{\nu}=g^{\mu\nu}\partial\mathbf{r/}%
x^{\nu}$, and at this point $\mathbf{r}$, $\mathbf{n=(}n_{x},n_{y}%
,n_{z}\mathbf{)}$ denotes the normal vector and $M\mathbf{n}$ symbolizes the
mean curvature vector field, a geometric invariant \cite{Liu1}. When first
seeing this expression (\ref{GM}) that apparently contains a term
$M\mathbf{n}$, many thinks that it has component along the normal direction
$\mathbf{n}$. It is\ in fact an operator exclusively defined on the tangent
plane to the surface at the given point for we have an operator relation
$\mathbf{p}\cdot\mathbf{n+n}\cdot\mathbf{p}=0$ with use of a relation$\ \nabla
_{2}\cdot\mathbf{n}=-2M$ \cite{diffgeom}. In flat space with $M=0$, operator
$\mathbf{p}\ $assumes its usual form $-i\hbar\nabla$. Noting that the mean
curvature $M$ is an extrinsic curvature, the momentum (\ref{GM}) actually
reflects a consequence of embedding the two-dimensional surface in
three-dimensional flat space. Other theoretical and experimental progresses
relating the quantum mechanics to the extrinsic curvature, please refer to
papers \cite{jk,dacosta,FC,Szameit,epl}. Why this momentum (\ref{GM}) is
advantageous over other canonical one comes from three respects: i) According
to Dirac, the commutator between the canonical momentum (\ref{GM}) and the
Hamiltonian of the system under consideration must take the same form as the
classical Poisson bracket for these two quantities, especially "\textit{for
systems for which the orders of the factors of products occurring in the
equations of motion are unimportant}" \cite{dirac}. As we showed in paper
\cite{Liu2} for the spherical surface, the canonical momentum $p_{\theta}$
violates this rule whereas the geometric one is completely compatible with it.
ii) From the point of the Dirac theory for systems with second class
constraints, the geometric momentum is nothing but the direct canonical
quantization of the tangential momentum on the surface \textit{provided that
the constraints are formulated in the phase-space rather than the
configuration one }\cite{Liu2,homma}. iii) For the two-dimensional sphere of
unit radius, the momentum (\ref{GM}) possesses a complete set of
eigenfunctions \cite{Liu2,Liu5} and have in principle observable consequence
\cite{Liu6}. In contrast, the canonical operator $p_{\theta}$ is not a
self-adjoint one.

Though Hall's formalism is quite generally applicable, e.g., easily to be
generalized for sphere in any dimensions \cite{H5}, while that of Kowalski and
Rembieli\'{n}ski has advantages to get the explicit form of the coherent
states on some compact manifolds, e.g., on the torus \cite{KR4}, the present
paper is limited within the two-dimensional sphere. The coherent states are
labeled by points in the associated classical phase space, namely the
cotangent bundle $T^{\ast}S^{2}=\left\{  \left.  (\mathbf{x},\mathbf{p}%
)\in\mathbb{R}^{3}\times\mathbb{R}^{3}\right\vert x^{2}=a^{2},\,\mathbf{x}%
\cdot\mathbf{p}=0\right\}  $ where $a$ is the radius of the sphere and both
position $\mathbf{x}$ and momentum$\ \mathbf{p}$ are defined for a single
particle on the sphere. In quantum mechanics, the relation $\,\mathbf{x}%
\cdot\mathbf{p}=0$ must be replaced by the operator identity$\ \mathbf{p}%
\cdot\mathbf{n+n}\cdot\mathbf{p}=0$.

In classical mechanics, the annihilation operators $\mathbf{a}$ on the sphere
are easily constructed via classical angular momentum $j$ together with
$\mathbf{x}$ and$\ \mathbf{p}$ as \cite{H5,KR3},%

\begin{equation}
\mathbf{a}=\cosh\left(  \frac{j}{m\omega r^{2}}\right)  \mathbf{x}%
+i\frac{r^{2}}{j}\sinh\left(  \frac{j}{m\omega r^{2}}\right)  \mathbf{p}%
.\label{AOC}%
\end{equation}
In quantum mechanics, with a "\textit{shifted, dimensionless angular
momentum"} \cite{H6} operator $S$ defined by,
\begin{equation}
S\equiv\sqrt{J^{2}/\hbar^{2}+1/4},\label{S}%
\end{equation}
with $\mathbf{J}$ being the angular momentum, the annihilation operators may
take the following dimensionless form \cite{H5,KR3},
\begin{equation}
\mathbf{Z}=\exp\left(  \frac{1}{2}\eta\right)  \left(  \left[  \cosh\left(
\eta S\right)  \mathbf{+}\frac{1}{2S}\sinh\left(  \eta S\right)  \right]
\frac{\mathbf{X}}{a}+\frac{1}{S}i\sinh\left(  \eta S\right)  \mathbf{P}\right)
\label{AO}%
\end{equation}
where $\mathbf{X}$ is the corresponding operator of Cartesian positions
$\mathbf{x}$ on the sphere, and $\mathbf{P\equiv J/}\hbar\mathbf{\times X/}a$
is a non-hermitian operator of quantity that looks like a momentum, and
$\delta\equiv1$ from Kowalski and Rembieli\'{n}ski \cite{KR3} and $\eta
\equiv\hbar/m\omega a^{2}$ from Hall \cite{H5} who also argued why this form
is necessary. The operator $S$ and the angular momentum possess simultaneous
eigenstates $\left\vert jm\right\rangle $, so the eigenvalues of $S$ are
simply,%
\begin{equation}
S\left\vert jm\right\rangle =(j+\frac{1}{2})\left\vert jm\right\rangle .
\end{equation}

\textit{First}, we are going to address the following issues: The direct
correspondence of $\mathbf{P}$ in (\ref{AO}) and $\mathbf{p}$ in (\ref{AOC})
is not accurate for we can easily show that the relation $\mathbf{P}%
\cdot\mathbf{n+n}\cdot\mathbf{P}=0$ does not hold. Thus we have to rewrite
$\mathbf{P}$ in (\ref{AO}) in terms of the geometric momentum (\ref{GM}) that
has exact classical correspondence.

After symmetrization of the $\mathbf{P}$, we have its corresponding Hermitian
operator $\mathbf{\Pi}$,%
\begin{equation}
\mathbf{\Pi}\equiv\frac{1}{2\hbar}\left(  \mathbf{J\times}\frac{\mathbf{X}}%
{a}-\frac{\mathbf{X}}{a}\mathbf{\times J}\right)  .\label{MM}%
\end{equation}
On the sphere parametrized by, ($0<\theta<\pi$, $0\leq\varphi\leq2\pi$)
\begin{equation}
x=a\sin\theta\cos\varphi,\text{ }y=a\sin\theta\sin\varphi,\text{ }%
z=a\cos\theta,
\end{equation}
three components of the operator $\mathbf{\Pi}$ are respectively,
\begin{align}
\Pi_{x} &  =-i(\cos\theta\cos\varphi\frac{\partial}{\partial\theta}-\frac
{\sin\varphi}{\sin\theta}\frac{\partial}{\partial\varphi}-\sin\theta
\cos\varphi),\label{deriop1}\\
\Pi_{y} &  =-i(\cos\theta\sin\varphi\frac{\partial}{\partial\theta}+\frac
{\cos\varphi}{\sin\theta}\frac{\partial}{\partial\varphi}-\sin\theta
\sin\varphi),\label{deriop2}\\
\Pi_{z} &  =-i(-\sin\theta\frac{\partial}{\partial\theta}-\cos\theta
).\label{deriop3}%
\end{align}
We have also,
\begin{equation}
\mathbf{\mathbf{\Pi}^{2}}\mathbf{=}\frac{J^{2}}{\hbar^{2}}+1.\label{rel3}%
\end{equation}
Rewriting $\mathbf{\Pi}$ as
\begin{equation}
\mathbf{p}=\frac{\hbar}{a}\mathbf{\Pi,}%
\end{equation}
we see that $\mathbf{\Pi}$ is nothing but the geometric momentum (\ref{GM}) on
the the two-dimensional sphere \cite{Liu1,Liu2,dm12}, despite a constant factor.

Noting that the ratio $\mathbf{X/}a$\ is the unit direction operator
$\mathbf{N}$,
\begin{equation}
\mathbf{N\equiv}\frac{\mathbf{X}}{a},\label{DO}%
\end{equation}
we see immediately how the annihilation operators (\ref{AO}) depend on $S$,
$\mathbf{N}$, and $\mathbf{p}$ as,
\begin{equation}
\mathbf{Z}=\exp\left(  \frac{1}{2}\eta\right)  \left(  \left[  \cosh\left(
\eta S\right)  \mathbf{-}\frac{1}{2S}\sinh\left(  \eta S\right)  \right]
\mathbf{N}+\frac{1}{S}i\sinh\left(  \eta S\right)  \mathbf{\Pi}\right)
.\label{AOGM}%
\end{equation}
With help of a series of relations, e.g.,%
\begin{align}
\mathbf{N\times\frac{\mathbf{J}}{\hbar}}\mathbf{=-\Pi+} &  i\mathbf{N,\frac
{\mathbf{J}}{\hbar}\times N=\Pi+}i\mathbf{N,}\label{rel1}\\
\mathbf{\Pi\times\frac{\mathbf{J}}{\hbar}}\mathbf{=N} &  \left(
\mathbf{\mathbf{\Pi}^{2}+}1\right)  \mathbf{,\frac{\mathbf{J}}{\hbar}\times
\Pi=}\left(  -\mathbf{\mathbf{\Pi}^{2}+}1\right)  \mathbf{N,}\label{rel2}%
\end{align}
the following properties of operator (\ref{AOGM}) can be easily proved. i) The
annihilation operators $\mathbf{Z}$ is a vector operator that satisfies,%
\begin{equation}
\lbrack J_{i}\text{, }Z_{j}]=i\varepsilon_{ijk}Z_{k}.\label{VO}%
\end{equation}
ii) The annihilation operators take following well-known form,
\begin{equation}
\mathbf{Z}=\alpha(J)\mathbf{N+}i\beta(J)\mathbf{\Pi=}\alpha(J)\frac
{\mathbf{X}}{a}\mathbf{\mathbf{+}}i\beta(J)\frac{a\mathbf{\mathbf{p}}}{\hbar
}\mathbf{,}\label{X+P}%
\end{equation}
where $\alpha(J)$ and $\beta(J)$ are two Hermitian operators solely depending
on the angular momentum $J$, and it formally coincides with the classical one
(\ref{AOC}) with considering the operator ordering.

\textit{Secondly}, we give the Schwinger boson representation of the
annihilation operators (\ref{X+P}).

The Schwinger bosons are two sets of annihilation and creation operators
$\vec{a}\equiv(a_{1},a_{2})$ and $\vec{a}^{\dagger}\equiv(a_{1}^{\dagger
},a_{2}^{\dagger})$ \cite{schwinger} which satisfy the bosonic commutation
relation $[a_{\alpha},a_{\beta}^{\dagger}]=\delta_{\alpha\beta}$ with
$\alpha,\beta=1,2$. The angular momentum operators are constructed out of
Schwinger bosons as: \cite{schwinger}
\begin{equation}
J_{k}\equiv a^{\dagger}\frac{\sigma_{k}}{2}a,\label{sch}%
\end{equation}
where $\sigma_{k}$ ($k=1,2,3$) denote the $k$-component of the Pauli matrices.
It is easy to check that the operators in (\ref{sch}) satisfy the $SU(2)$ Lie
algebra with the $SU(2)$ Casimir operator:
\begin{equation}
J^{2}\equiv{\frac{\vec{a}^{\dagger}\cdot\vec{a}}{2}}\left(  {\frac{\vec
{a}^{\dagger}\cdot\vec{a}}{2}}+1\right)  .
\end{equation}
Thus the angular momentum quantum number and operator $S$ (\ref{S}) can be
expressed by the eigenvalues of total occupation number operator $n\equiv
n_{1}+n_{2}$,
\begin{equation}
j={\frac{\left(  n_{1}+n_{2}\right)  }{2}}\equiv{\frac{n}{2},}\text{ {and }%
}{S=\frac{n+1}{2},}%
\end{equation}
where the occupation numbers are respectively $n_{1}=a_{1}^{\dagger}a_{1}$ and
$n_{2}=a_{2}^{\dagger}a_{2}$, and the eigenvalues of the occupation number
operator are $n_{\alpha}=0,1,2,...$. The unit direction vector $\mathbf{N}$
and the geometric momentum $\mathbf{\Pi}$ are respectively \cite{Liu1,GK,Liu4}%
,
\begin{align}
\mathbf{N} &  \mathbf{=}\frac{1}{2}\left(  \frac{1}{\sqrt{S(S+1)}}%
\mathbf{A}+\mathbf{A}^{\dagger}\frac{1}{\sqrt{S(S+1)}}\right)  ,\label{NSchw}%
\\
\mathbf{\Pi} &  \mathbf{=}\frac{1}{2}\left(  \sqrt{S}\mathbf{B}\frac{1}%
{\sqrt{S}}+\frac{1}{\sqrt{S}}\mathbf{B}\sqrt{S}\right)  ,\label{PiSchw}%
\end{align}
where,%
\begin{align}
\mathbf{A} &  \mathbf{=}\left(  \frac{1}{2}\left(  a_{2}a_{2}-a_{1}%
a_{1}\right)  ,-\frac{i}{2}\left(  a_{2}a_{2}+a_{1}a_{1}\right)  ,a_{2}%
a_{1}\right)  ,\\
\mathbf{B} &  =\frac{i}{2}\left(  \mathbf{A}^{\dagger}-\mathbf{A}\right)  .
\end{align}
Thus, we have the Schwinger boson representation of the annihilation operators
(\ref{AOGM})
\begin{align}
Z_{+} &  \equiv Z_{x}+iZ_{y}=\frac{\exp\left(  1/2\right)  }{2}\left(
\frac{\exp\left(  S\right)  }{\sqrt{S(S+1)}}a_{2}a_{2}-\frac{\exp\left(
-S\right)  }{\sqrt{S(S-1)}}a_{1}^{\dag}a_{1}^{\dag}\right)  ,\label{Z+}\\
Z_{-} &  \equiv Z_{x}-iZ_{y}=\frac{\exp\left(  1/2\right)  }{2}\left(
-\frac{\exp\left(  S\right)  }{\sqrt{S(S+1)}}a_{1}a_{1}+\frac{\exp\left(
-S\right)  }{\sqrt{S(S-1)}}a_{2}^{\dag}a_{2}^{\dag}\right)  ,\label{Z-}\\
Z_{z} &  =\frac{\exp\left(  1/2\right)  }{2}\left(  \frac{\exp\left(
S\right)  }{\sqrt{S(S+1)}}a_{2}a_{1}+\frac{\exp\left(  -S\right)  }%
{\sqrt{S(S-1)}}a_{2}^{\dag}a_{1}^{\dag}\right)  .\label{Z0}%
\end{align}
From these operators (\ref{Z+})-(\ref{Z0}), we can easily verify the following
relations:%
\begin{equation}
\lbrack Z_{i},Z_{j}]=0,\text{ }Z^{2}=1.
\end{equation}

To conclude and remark, we see that\textit{\ }the annihilation operators whose
eigenstates are coherent states on a two-dimensional sphere turn out to be of
the expected form $\alpha\mathbf{x}+i\beta\mathbf{p}$, where $\alpha$, $\beta$
and $\mathbf{x}$, $\mathbf{p}$ are four Hermitian operators, and as expected,
the annihilation operators are expressible in the Schwinger boson
representation. Moreover, we make a general conjecture that on an arbitrary
surface the coherent states generated by the annihilation operators, once they
exist, the similar relation (\ref{X+P}) holds true between the annihilation
operators and the geometric momentum. For two-dimensional surfaces, the mean
curvature is an extrinsic one while the gaussian curvature is an intrinsic
geometric property. So, the geometric momentum is not obtainable from the
intrinsic point of view. Since the geometric momentum is manifestly a
consequence of embedding the two-dimensional sphere in the three-dimensional
flat space, the coherent states reflects some aspects of the extrinsic
geometric properties of the surfaces.

\begin{acknowledgments}
This work is financially supported by National Natural Science Foundation of
China under Grant No. 11175063.
\end{acknowledgments}

\end{document}